\documentclass[prd,twocolumn,amsmath,amssymb,floatfix,superscriptaddress,nofootinbib,preprintnumbers]
{revtex4-1}
\usepackage{float}
\usepackage{subcaption}
 \usepackage[utf8]{inputenc}
\usepackage{graphicx}
\usepackage{amssymb}
\usepackage{amsmath}
\usepackage{bm}
\usepackage{color}
\usepackage{enumitem}
\usepackage{booktabs}
\usepackage{caption}

\DeclareMathAlphabet\mathbfcal{OMS}{cmsy}{b}{n}
\definecolor{darkgreen}{RGB}{50,150,0}
\definecolor{purple}{cmyk}{0.5,0.75,0,0}
\definecolor{darkpurple}{RGB}{128,0,128}
\definecolor{ultramarine}{rgb}{0.07, 0.04, 0.56}
\definecolor{cadmiumgreen}{rgb}{0.0, 0.42, 0.24}
\definecolor{indigo(dye)}{rgb}{0.0, 0.25, 0.42}

\usepackage[linktocpage=true]{hyperref}
\hypersetup{
colorlinks=true,
citecolor=ultramarine,
linkcolor=cadmiumgreen,
urlcolor=indigo(dye),
pdfauthor={},
pdftitle={},
pdfsubject={}
}

\def\aj{{\rm AJ}}                   
\def\apj{{\rm ApJ}}                 
\def\aap{{\rm A\&A}}                
\def\mnras{{\rm MNRAS}}             
\def\prd{{\rm Phys.~Rev.~D}}        

\def\mnras{Mon.\ Not.\ R.\ Astron.\ Soc.\ }

\begin{document}
\preprint{YITP-SB-16-18}

\title{Observable Predictions for Massive-Neutrino Cosmologies with Model-Independent Dark Energy}

\author{Ana Diaz Rivero}
\email{adiazrivero@g.harvard.edu}
\affiliation{Department of Physics, Harvard University, Cambridge, MA 02138, USA}

\author{V. Miranda}
\email{vivianmiranda@email.arizona.edu }
\affiliation{Steward Observatory, Department of Astronomy, University of Arizona, Tucson, Arizona, 85721, USA}
\affiliation{Center for Particle Cosmology, Department of Physics and Astronomy,\\University of Pennsylvania, Philadelphia, Pennsylvania 19104, USA}

\author{Cora Dvorkin}
\email{cdvorkin@g.harvard.edu}
\affiliation{Department of Physics, Harvard University, Cambridge, MA 02138, USA}

\begin{abstract}
\noindent We investigate the bounds on the sum of neutrino masses in a cosmic-acceleration scenario where the equation of state $w(z)$ of dark energy (DE) is constructed in a model-independent way, using a basis of principal components (PCs) that are allowed to cross the phantom barrier $w(z)=-1$. We find that the additional freedom provided to $w(z)$ means the DE can undo changes in the background expansion induced by massive neutrinos at low redshifts. This has two significant consequences: (1) it leads to a substantial increase in the upper bound for the sum of the neutrino masses ($M_{\nu} < 0.33 - 0.55$ eV at the 95\% C.L. depending on the data sets and number of PCs included) compared to studies that choose a specific parametrization for $w(z)$; and (2) it causes $\sim1\sigma$ deviations from $\Lambda$CDM in the luminosity distance and the Hubble expansion rate at higher redshifts ($z \gtrsim 2$), where the contribution of DE is subdominant and there is little constraining data. The second point consequently means that there are also observable deviations in the shear power spectrum and in the matter power spectrum at low redshift, since the clustering of matter throughout cosmic time depends on the expansion rate. This provides a compelling case to pursue high-$z$ BAO and SN measurements as a way of disentangling the effects of neutrinos and dark energy. Finally, we find that the additional freedom given to the dark energy component has the effect of lowering $S_8$ with respect to $\Lambda$CDM.
\end{abstract}

\maketitle

\section{Introduction}

The source of the observed accelerated expansion of the universe ({\it dark energy} from now on) has remained an elusive component since its discovery, two decades ago  \cite{Riess:1998cb,Perlmutter:1998np}. The standard paradigm in cosmology, the $\Lambda$CDM model, assumes that the cosmological constant $\Lambda$ is the source of acceleration, and its six free parameters are constrained to exquisite precision by current Cosmic Microwave Background (CMB; e.g. \cite{Aghanim:2018eyx}), Baryon Acoustic Oscillations (BAO; e.g. \cite{2017MNRAS.470.2617A}) and weak lensing (WL) measurements (e.g. \cite{Hikage:2018qbn}). Despite the success of the $\Lambda$CDM model in providing a solid statistical fit to these various probes, no satisfying theoretical model elucidates the microphysical origin of the cosmological constant.

Cosmological probes have become precise enough that they can be used to look for physics beyond the Standard Model of particle physics; for example, to study neutrino properties. In the early universe, neutrinos are relativistic, while at late times they become non-relativistic, with their mass constituting a non-negligible fraction of the total dark matter component. The conversion of radiation to hot dark matter affects the Hubble expansion, and the residual streaming velocities are still significant enough at low redshifts to slow down the growth of structure. This means that neutrinos affect cosmology at both the background and perturbation level. 

Direct detection experiments on Earth have measured the three mixing angles and the two mass-squared splittings of the three neutrino mass eigenstates with high precision \cite{deSalas:2017kay}. However, so far cosmology appears to be the most sensitive probe to the absolute mass scale of neutrinos. To date, the most stringent upper limit of the sum of neutrino masses ($M_{\nu} \equiv \sum m_{\nu} < 0.12$ eV) is given by the combination of CMB TT,TE,EE power spectra and lensing from \textit{Planck}, together with BAO data, assuming the standard $\Lambda$CDM scenario \cite{Aghanim:2018eyx}.

When probing the presence of additional light particles (such as neutrinos) with cosmological data sets, dark energy acts as a source of systematic uncertainty that needs to be marginalized over. Extensions to the base $\Lambda$CDM model in terms of dark energy are generally implemented by allowing the equation of state $w$ to be a function of redshift $z$. The most common parametrization for $w(z)$ is known as the Chevallier-Polarski-Linder (CPL) parametrization, given by $w(z) = w_0 + w_a z/(1+z)$, i.e. it has two free parameters $\{w_0,w_a\}$. However, the fact that there is no firm theoretical backbone to support any particular parametrization for $w(z)$ motivates the study of the effects of dark energy using non-parametric methods, such as Principal Components Analysis (PCA). 

We consider a dark energy scenario known as the Smooth Dark Energy Paradigm~\cite{Mortonson:2008qy}, which makes a set of assumptions about the microphysical nature of dark energy. It assumes that the source of dark energy: (1) does not cluster inside the horizon, (2) interacts only gravitationally with dark matter and baryons, and that (3) gravity is set by General Relativity. An analysis of such a scenario has been performed in the past, including cross-checks between the background expansion of the universe and the growth of linear perturbations \cite{Ruiz:2014hma}, as a smoking gun to falsify models where dark energy follows these assumptions \cite{Miranda:2017mnw}. In fact this paper partially generalizes Ref.~\cite{Miranda:2017mnw} with the inclusion of the sum of neutrino masses, except that in that work 20 principal components (PCs) were used, and here we restrict ourselves to considering 1, 3 or 5 PCs. Indeed, no more than five principal components are necessary to assess the degeneracies between the sum of neutrino masses and dark energy properties that may not be apparent in specific parametrizations of $w(z)$. At the same time such generalizations do not need to have signal-to-noise in very high modes: the case of three principal components already shows order unity changes on $M_\nu$ constraints.

There are two distinct regimes when studying extensions to $\Lambda$CDM concerning the dark energy equation of state. In the first, and most commonly considered regime, the equation of state has a lower bound set by the value of the cosmological constant, i.e. $w(z) \geq -1$.  Alternatively, models that do not satisfy this criterion are in the ``phantom dark energy" regime. Ref.~\cite{Vagnozzi:2018jhn} showed that the bound on the sum of the neutrino masses is more stringent with a CPL non-phantom dark energy source than in a standard $\Lambda$CDM cosmology. Indeed, the expansion rate is faster for non-phantom dynamical dark energy than for $\Lambda$CDM. The sum of neutrino masses is consequently pushed downwards to keep the angular diameter distance to the surface of the last scattering fixed. Conversely, in phantom dark energy scenarios the opposite is true and thus the bounds on $M_{\nu}$ degrade. Note that the general argument is independent of the chosen CPL parametrization for the dark energy equation of state. 

In this work, we aim to generalize these findings by studying constraints on the sum of neutrino masses in a cosmology with a model-independent, time-varying dark energy component, without forbidding the cross into the phantom regime. In Section \ref{sec:data} we introduce our methodology and the data sets used. In Section \ref{sec:main_analysis} we present our main results, and in Section \ref{sec:discussion} we discuss the implications of our findings.


\section{Data and Methods}
\label{sec:data}

\begin{figure}
\includegraphics[width=0.5\textwidth]{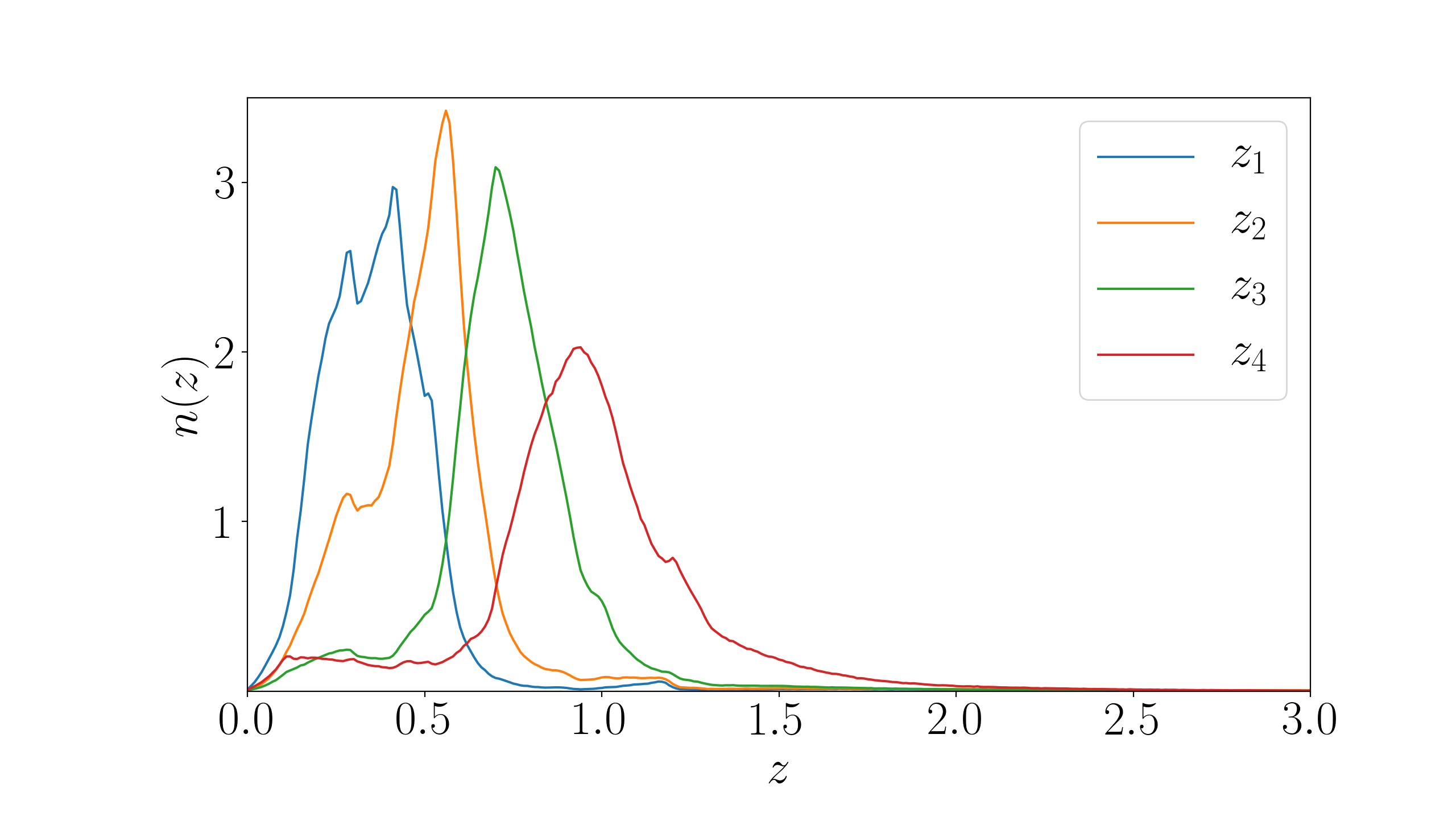}
\caption{Tomographic distribution of lenses $n(z)$ in the Dark Energy Survey\cite{Krause:2017ekm}.}\label{fig:nz}
\end{figure}

The data sets considered in this work include: BAO data from BOSS DR12 \cite{2016MNRAS.460.4210G}, 6dF Galaxy Survey (6dF) \cite{2011MNRAS.416.3017B} and SDSS DR7 Main Galaxy Sample (MGS) \cite{d5dd3c2b697c4bf68f1f8f19843253ab}); 
the full 2015 lensed \textit{Planck} CMB temperature and polarization data \cite{Aghanim:2015wva} and lensing reconstruction \cite{Ade:2015zua}; Dark Energy Survey (DES) four-bin tomographic weak lensing data \cite{Abbott:2017wau,Krause:2017ekm}, shown in Figure \ref{fig:nz}; and the Pantheon supernovae (SN) sample \cite{Scolnic:2017caz}. The latter covers a redshift range $0.01 < z < 2.3$, and the BAO measurements lie within this range. To account for the non-linear scales in the matter power spectrum we adopt the HALOFIT fit \cite{Takahashi:2012em}, and a mapping between arbitrary $w(z)$ and a constant dark energy equation of state is implemented as in Ref. \cite{Casarini:2016ysv}.

We perform a Markov Chain Monte Carlo (MCMC) likelihood analysis with a modified version of the CosmoMC code \cite{Lewis:2013hha,Lewis:2002ah,Lewis:1999bs,Howlett:2012mh}. 
We ran two different sets of chains. The first of these, which we will refer to as the \textit{All} chains, includes all the aforementioned data sets. The second one, which we refer to as the \textit{Reduced} chains, does not include the DES weak lensing. The data sets used for each of these chains are summarized in Table \ref{tab:data sets}. Note that there is a known (2.4$\sigma$) tension between the value of $S_8$ from \textit{Planck} ($0.848^{+0.024}_{-0.024}$) and DES ($0.782^{+0.027}_{-0.027}$) \cite{Troxel:2017xyo}, and we will discuss this in Sections \ref{sec:main_analysis} and \ref{sec:discussion}. Nevertheless, note that CMB lensing reconstruction (included in both data sets) favors a lower $S_8$ that than inferred from the CMB temperature and polarization under the $\Lambda$CDM model \cite{Ade:2015xua}, and thus more in accordance with the value of $S_8$ measured with WL.

\begin{widetext}

\begin{minipage}{\linewidth}
  \footnotesize
    \bigskip
 \centering
 \captionof{table}{Data sets} 
 \begin{tabular}{ | c | c | c |}
\hline
  & \textit{All}  & \textit{Reduced}   \\ \hline
CMB & Full \textit{Planck} 2015   & Full \textit{Planck} 2015  \\ [0.2cm]
BAO & BOSS DR12 + 6dF + MGS & BOSS DR12 + 6dF + MGS \\ [0.2cm]
SN & Pantheon & Pantheon \\ [0.2cm]
WL & DES & - \\ [0.2cm]
\hline
\end{tabular}\par \label{tab:data sets}
Data sets that define the \textit{All} and \textit{Reduced} sets of chains.
\end{minipage}

\end{widetext}


We represent the dark energy equation of state as
\begin{align}\label{eq:wz_PC}
w(z) = w_\text{fiducial} + \sum_{i=1}^{N_\text{PC}} \alpha_i e_i(z),
\end{align}
where $e_i(z)$, with $i=1,...,N_\text{PC}$, are the principal components of a covariance matrix of perturbations around the fiducial model $w_\text{fiducial} = -1$. The principal components used here have support in the range $0 < z < z_\text{max}$, and for $z > z_{\rm max}$ we extrapolate assuming $w = -1$. 

We construct the PC basis from the eigenvectors of the Wide Field Infrared Survey Telescope (WFIRST) experiment \cite{2017arXiv170201747H} supernovae Fisher matrix, which has SN data up to $z_{\rm max} = 3$ \cite{2018ApJ...867...23H}. Note that we do not include simulated WFIRST data in our likelihood analysis; instead, our goal is to use current datasets to make predictions for future experiments, such as WFIRST, that will gather SN and WL data. Finally, a \textit{Planck}-like likelihood is also added to the total Fisher matrix. The shape and priors of the PCs, as well as details on how the basis was constructed, can be found in Ref.~\cite{Miranda:2017mnw}. For this work we ran three different sets of chains for each of the \textit{All} and \textit{Reduced} data sets with varying numbers of principal components, $N_{\rm PC} = \{1,3,5\}$.

 Note that the minimum supernova redshift from the Pantheon sample imposes $z_{\rm min} = 0.01$, which means that in fact for $0 < z < z_{\rm min}$ the PCs can oscillate significantly. 
 There is no fundamental reason why $w(z)$ must not change arbitrarily at ultra-low redshift~\cite{2009PhRvD..80f7301M}, and the PCs capture that possibility.

The vector of model parameters for the chains is given by
\begin{align}
\vec{\theta} &=\{ \Omega_c, \theta_A, \Omega_b, \tau, n_s, \ln A_s, \tau, \alpha_1,..., \alpha_{N_\text{PC}}, M_{\nu} \}.
\end{align}
Here, $\Omega_c$ is the cold dark matter density, $\theta_A$ is the angular size of the horizon at the time of recombination, $\Omega_b$ is the baryon density, $\tau$ is the reionization optical depth, and $A_s$ and $n_s$ are the initial curvature power spectrum amplitude and tilt. We define $M_{\nu} \equiv \sum m_{\nu}$ as the sum of the neutrino masses and assume the so-called degenerate hierarchy, where all three neutrino eigenstates are equally massive (i.e. the mass of the $i$th neutrino eigenstate is $m_{\nu,i} = M_{\nu}/3$ for $i = \{1,2,3\}$). Finally, the $\alpha_i$ parameters are the amplitudes of the PCs $e_i(z)$, as per Eq. \eqref{eq:wz_PC}. 

To evaluate the joint effect of massive neutrinos and a model-independent equation of state for dark energy, we consider a variety of cosmological probes of geometry (the Hubble expansion rate as a function of redshift $H(z)$ and the luminosity distance $D_L(z)$) and probes of the growth (the matter power spectrum $P(k)$, the shear power spectrum $P_{\ell}^{\kappa}$, and $S_8 \equiv \sigma_8 \sqrt{\Omega_m/0.3}$, where $\sigma_8$ is the amplitude of the linear matter power spectrum at a scale of $8h^{-1}$ Mpc\footnote{Generally, the parameter combination $f(z)\sigma_8$ (where $f(z)$ is the logarithmic derivative of the linear growth rate of matter fluctuations) is of interest to measure the growth of structure because it is insensitive to galaxy bias. In the $\Lambda$CDM cosmology, $f(z)$ is well approximated by $f(z) = \Omega_{\rm m}(z)^{0.545}$. Different conventions exist for the definition of $S_8$ however: the value of the exponent varies and sometimes $\Omega_m$ is divided by a fiducial value $\Omega_{\rm m,fid}$. Here we choose 0.5 as our exponent and use $\Omega_{m,{\rm fid}} = 0.3$ for a direct comparison of $S_8$ with values quoted in Ref.~\cite{Hikage:2018qbn}.}).

Since conventions for the calculation of the shear power spectrum vary across the literature, we briefly outline the procedure used in this paper. The shear power spectrum is given by
\begin{align}\label{eq:plkappa}
P_{\ell}^{\kappa} &= \frac{2\pi^2}{\ell^3} \int_0^{z_s} dz \frac{W^2(z) \chi(z)}{H(z)} \Delta^2\left(k,z\right) \\
&= \frac{2\pi^2}{\ell^3} \int_0^{z_s} dz \; F(z) \; \Delta^2\left(k,z\right),
\end{align}
where $\Delta^2$ is the dimensionless matter power spectrum,
\begin{equation}
\Delta^2(k,z) = \frac{k^3 P(k,z)}{2 \pi^2}
\end{equation}
and
\begin{equation}
F(z) = \frac{W^2(z)\chi(z)}{H(z)} \frac{1}{c^3}
\end{equation}
is known as the lensing weight function. Note that we use the Limber approximation \cite{1953ApJ...117..134L}, whereby $k\approx \ell/\chi(z)$, to evaluate the integral.

For a dark energy model that has an equation of state that is a function of redshift (and a flat universe),
\begin{multline}
H(z) = H_0 \Big[\Omega_m(1+z)^3 + \Omega_{\nu}(z) + \\
\Omega_{\rm DE} \times \text{exp}\big[3 \int_0^z d \ln(1+z')(1+w(z'))\big] \Big]^{1/2},
\end{multline}
where $\Omega_{\nu}(z)h^2 \approx M_{\nu}(1+z)^3/93.14$ eV when neutrinos become non-relativistic, $\chi$ is the comoving distance:
\begin{equation}
\chi(z) = c \int dz \frac{1}{H(z)},
\end{equation}
and 
$W(z)$ is the weight function,
\begin{equation}
W(z) = \frac{3}{2} \Omega_{m}H_0^2 g(z) (1+z).
\end{equation}
Here $g(z)$ is known as the efficiency factor, and it is defined as
\begin{equation}
g(z) \equiv \chi(z) \int_z^{\infty} dz' n(z') \frac{\chi(z')-\chi(z)}{\chi(z')},
\end{equation}
where $n(z)$ is the distribution of lenses (normalized such that $\int n(z) dz = 1$). The four tomographic bins for DES \cite{Krause:2017ekm} are shown in Figure \ref{fig:nz}.

Throughout the remainder of the paper, we refer to the set of chains that vary neutrino mass and $w(z)$ constructed with PCs as $\nu w$CDM. The chains that vary neutrino mass but $w= -1$ are referred to as $\nu\Lambda$CDM, and those with the usual cosmology (i.e. with $w=-1$ and $M_{\nu} = 0.06$ eV) are referred to as $\Lambda$CDM.


\section{Results}\label{sec:main_analysis}

\begin{figure*}
	\centering
	\begin{subfigure}[t]{0.497\textwidth}{\includegraphics[width=0.9\textwidth]{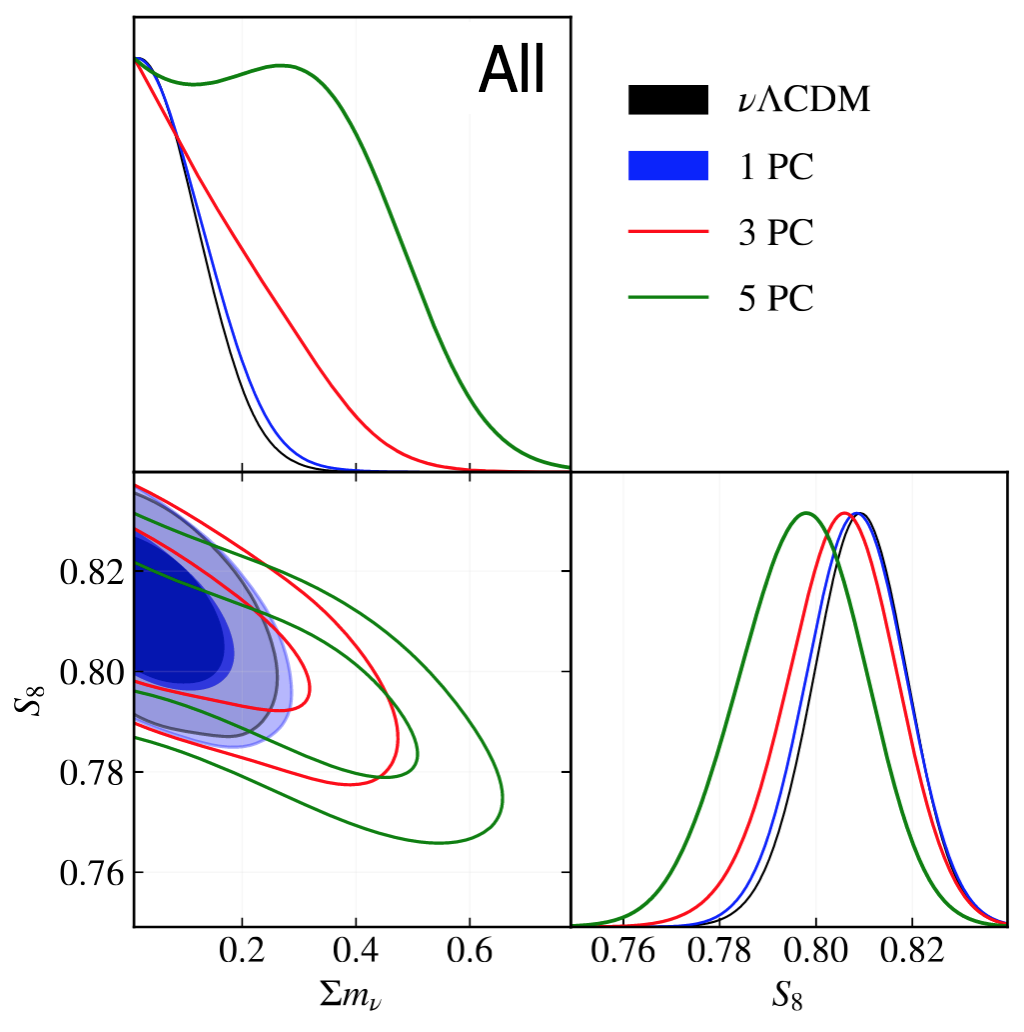}}	
		\caption{}
	\end{subfigure}
	\begin{subfigure}[t]{0.497\textwidth}{\includegraphics[width=0.9\textwidth]{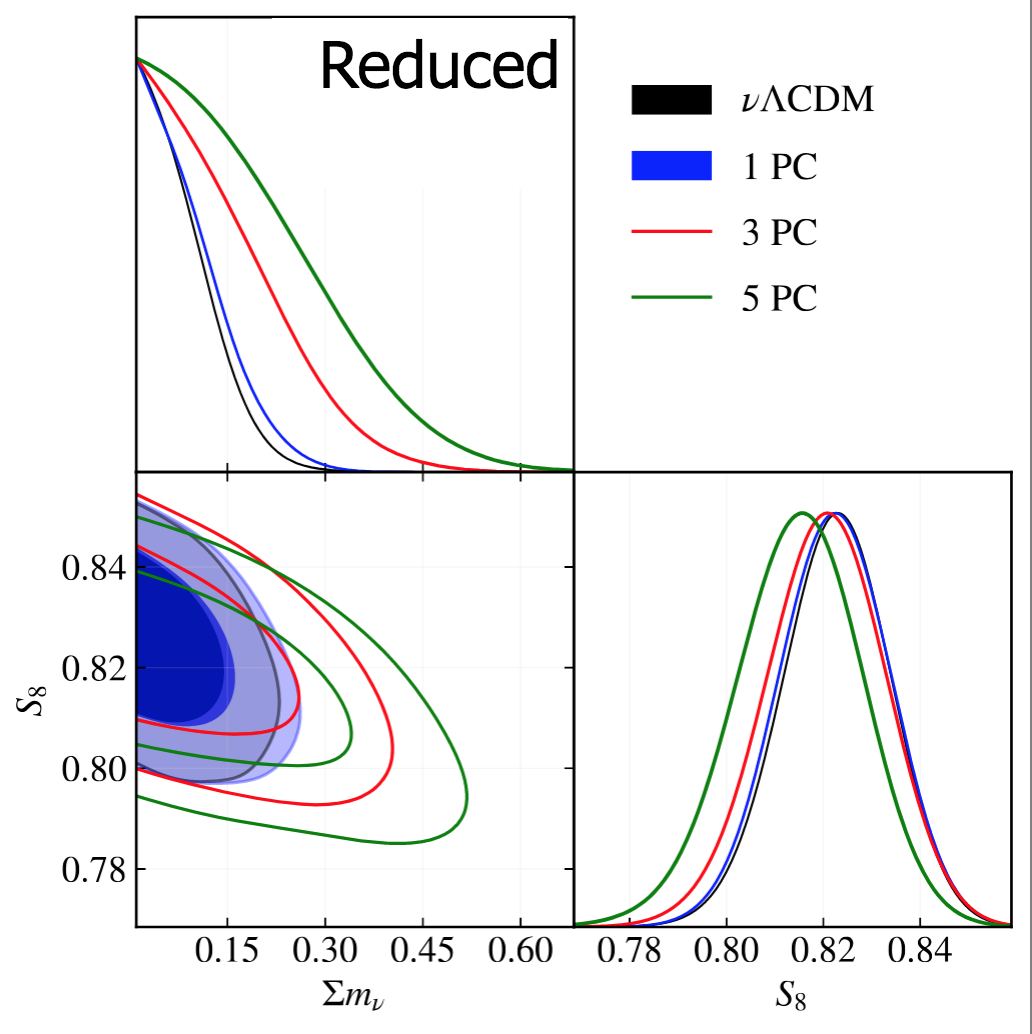}}
		\caption{}
	\end{subfigure}
	\\
\caption{Constraints on the sum of neutrino masses and $S_8 \equiv \sigma_8 \sqrt{\Omega_m/0.3}$ from the $\nu w$CDM and $\nu\Lambda$CDM scenarios. (a) \textit{All} data set. (b) \textit{Reduced} data set.}\label{fig:mnu_S8}
\end{figure*}

\begin{figure*}[t!]
\hspace*{-1cm} 
\includegraphics[width=1.15\textwidth]{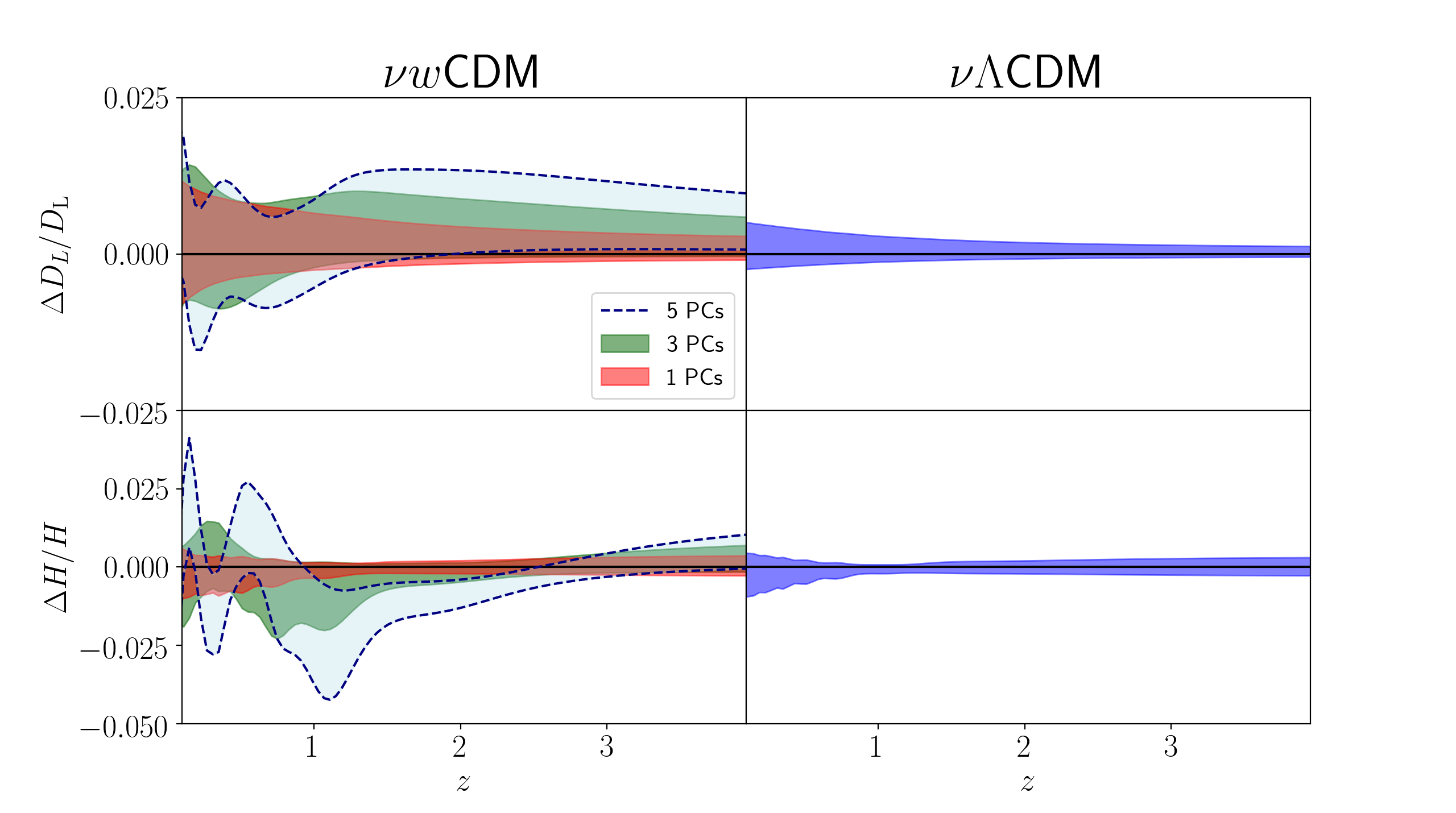}
\caption{Fractional difference for the luminosity distance $D_L(z)$ (top row) and the Hubble expansion rate as a function of redshift $H(z)$ (bottom row) for $\nu w$CDM (left) and $\nu\Lambda$CDM (right), with respect to $\Lambda$CDM, for the \textit{All} data set. Bands correspond to $1\sigma$ confidence levels.
}\label{fig:Hz_DL}
\end{figure*}

Tables \ref{tab:posteriors_nuwcdm} and \ref{tab:posteriors_nulcdm} show the best-fit and 95\% C.L. bounds on several cosmological parameters, including $M_{\nu}$ and $S_8$, marginalized over principal component amplitudes, for both models and data sets. For a given data set combination, the posteriors in  the $\nu w$CDM chains are significantly degraded with respect to $\Lambda$CDM, which is expected since we are marginalizing over several more parameters\footnote{Ref.~\cite{Vagnozzi:2018jhn} points out that since the expansion rate in models where the dark energy is exclusively phantom  is higher than that of the cosmological constant, this degrades constraints on $M_{\nu}$ to maintain the distance to the last scattering surface fixed (and the converse is true in non-phantom dark energy models, despite the larger parameter space). Note, however, that while this contributes to our wider posteriors to some extent, our dark energy is not forced to be phantom $-$ and indeed Figure \ref{fig:Hz_DL} shows that the fractional difference between $H(z)_{\nu w \rm{CDM}}$ and $H(z)_{\Lambda\rm{CDM}}$ is not always positive.}. More important, however, is the fact that the extra freedom given to the dark energy in the $\nu w$CDM chains is able to undo some of the changes induced by neutrinos, which consequently means that larger neutrino masses can be accommodated within the data: notice the difference in the $\nu w$CDM posteriors shown in Figure \ref{fig:mnu_S8}. As we increase the number of PCs - and thus give the dark energy more freedom - the allowed neutrino masses increase considerably, with an upper bound of $\lesssim 0.38$ $(0.33)$ eV (95\% C.L.) with 3 PCs and $\lesssim 0.55$ $(0.42)$ eV (95\% C.L.) with 5 PCs in the \textit{All} (\textit{Reduced}) data set. Compare this to the results in Ref. \cite{Vagnozzi:2018jhn}, where the allowed parameter space for massive neutrinos in a cosmology with a phantom, CPL-parametrized $w(z)$ was (slightly) broader than $\Lambda$CDM but still quite limited, with $M_{\nu} < 0.19$ eV. Clearly, letting the behavior of dark energy be dictated by the data instead of imposing a specific parametrization can significantly open up the allowed parameter space for $M_{\nu}$. Furthermore, comparing the two different data set combinations for a given model reveals that low-redshift growth information has a slight preference for cosmologies with massive neutrinos: for both the $\nu\Lambda$CDM and $\nu w$CDM chains, $M_{\nu}$ is pushed upwards when going from the \textit{Reduced} to the \textit{All} data set. 
This makes sense considering that WL surveys, such as DES, tend to have a valus of $S_8$ that are $2-3\sigma$  lower than that of \textit{Planck} \cite{Hikage:2018qbn}, and $S_8$ and $M_{\nu}$ are anti-correlated. Note that this anti-correlation is also why in going from $\nu \Lambda$CDM to $\nu$wCDM the value of $S_8$ decreases. 

\begin{widetext}

\begin{minipage}{\linewidth}
  \footnotesize
    \bigskip
 \centering
  \captionof{table}{$\nu w$CDM} 
 \begin{tabular}{|c|c|c|c|c|c|c|}
\hline
& \multicolumn{3}{c|}{\textit{All}} & \multicolumn{3}{c|}{\textit{Reduced}} \\ \hline
$N_{\rm PC}$ & 1 & 3 & 5 & 1 & 3 & 5 \\ \hline
$M_{\nu}$ [eV] & $< 0.23^{}_{}$ & $< 0.38^{}_{}$ & $< 0.55^{}_{}$
& $< 0.21^{}_{}$ & $< 0.33^{}_{}$ & $< 0.42^{}_{}$ \\[0.2cm]
$S_8$ & $0.81^{+0.02}_{-0.02}$ & $0.81^{+0.02}_{-0.02}$ & $0.80^{+0.02}_{-0.03}$
& $0.82^{+0.02}_{-0.02}$ & $0.82^{+0.02}_{-0.03}$ & $0.82^{+0.03}_{-0.03}$\\[0.2cm]
\hline
\end{tabular} \par \label{tab:posteriors_nuwcdm}
Mean and 95\% C.L. errors for $M_{\nu}$ and $S_8$ in the $\nu w$CDM chains, for the the \textit{All} and \textit{Reduced} group of chains, and varying number of PCs. Entries with no subscript correspond to chains where only an upper bound was obtained for that parameter.
\end{minipage}

\bigskip

\begin{minipage}{\linewidth}
  \footnotesize
    \bigskip
 \centering
 \captionof{table}{$\nu\Lambda$CDM} 
 \begin{tabular}{ | c | c | c |}
\hline
& \textit{All} & \textit{Reduced } \\ \hline
$M_{\nu}$ [eV] & $< 0.21^{}_{}$ & $< 0.20^{}_{}$ \\[0.2cm]
$S_8$ & $0.81^{+0.02}_{-0.02}$ & $0.82^{+0.02}_{-0.02}$ \\[0.2cm]
\hline
\end{tabular}\par \label{tab:posteriors_nulcdm}
Mean and 95\% C.L. errors for $M_{\nu}$ and $S_8$ in the $\nu\Lambda$CDM chains, for the the \textit{All} and \textit{Reduced} group of chains. Entries with no subscript correspond to chains where only an upper bound was obtained for that parameter.
\end{minipage}

\end{widetext}

\begin{figure*}[t!]
\hspace*{-1cm} 
\includegraphics[width=1.15\textwidth]{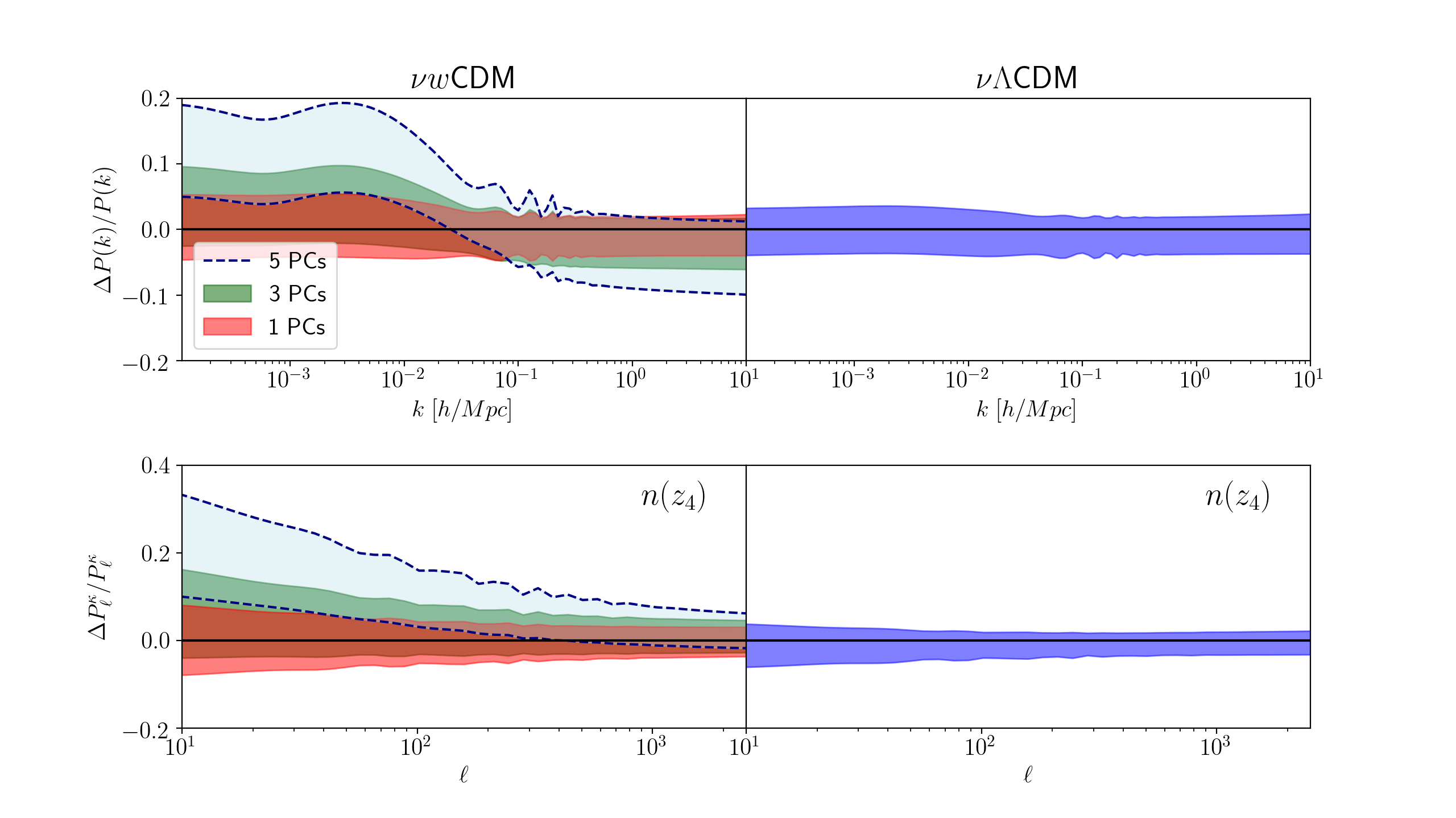}
\caption{Fractional difference for the matter power spectrum $P(k)$ at $z=0$ (top row) and the shear power spectrum $P_l^{\kappa}$ (bottom row) for $\nu w$CDM (left) and $\nu\Lambda$CDM (right) with respect to $\Lambda$CDM, for the \textit{All} data set. Bands correspond to $1\sigma$ confidence levels.}\label{fig:power_spectra}
\end{figure*}

\newpage

For the remainder of the paper we focus on the chains with the \textit{All} data set; we found that the differences in the remaining posteriors between the \textit{All} and the \textit{Reduced} data sets were marginal, so to avoid clutter we choose to show results for the more comprehensive of the two. For reference, Appendix \ref{app:A} shows the results for the \textit{Reduced} data set.

Figure \ref{fig:Hz_DL} shows the fractional difference of $\nu w$CDM (left) and $\nu\Lambda$CDM (right) with respect to $\Lambda$CDM for the luminosity distance $D_L(z)$ (top row) and the Hubble expansion rate as a function of redshift $H(z)$ (bottom row), for the chains with 1, 3, and 5 PCs. Note that the bands shown correspond to the $1\sigma$ confidence levels. Looking at the top left panel of Figure \ref{fig:Hz_DL}, it can be seen that the luminosity distance in the $\nu w$CDM and $\Lambda$CDM models agree at small redshifts ($z \lesssim 1$), but at higher redshifts there is an amplitude difference between them, reaching a $\sim 1 \sigma$ difference for the case with 3 PCs, and $>1 \sigma$ with 5 PCs. This is indicative of the fact that neutrinos change the background evolution, and while at low redshifts the dynamical dark energy can counteract these changes, the same is not true for redshifts $z \gtrsim 2$, where it becomes subdominant. Furthermore, we do not have support from BAO and SN data in this redshift range, so we find that at these redshifts the changes massive neutrinos induce in the background expansion are disfavored by the $\Lambda$CDM model. The fact that at low redshifts the dark energy is able to counteract the effects induced by neutrinos is what allows for the much larger values of $M_{\nu}$ in $\nu w$CDM than in $\Lambda$CDM, as shown in Figure \ref{fig:mnu_S8}.

This effect is also visible in the posterior for $H(z)$ (naturally, since $H(z)$ and $D_{\rm L}(z)$ are related by an integral). Between $z \approx 1 - 2.5$, the expansion rate in the $\nu w$CDM cosmology is lower than in $\Lambda$CDM, which is indicative of the dark energy behaving as a phantom component (indeed we see the $w(z)$ posteriors having values lower than $-1$ in this range, consistent with what is found in Ref. \cite{Zhao:2017cud}). This behavior is driven by the dark energy attempting to appease the aforementioned ``discrepancy" in the background favored by BAO data versus massive neutrinos.

Figure \ref{fig:power_spectra} shows the fractional difference in the matter power spectrum (at $z=0$) and the cosmic shear power spectrum for the same combination of models as the ones used in Figure \ref{fig:Hz_DL}. On small scales we can see the characteristic suppression of power on scales below the free-streaming length of neutrinos ($\mathcal{O}(10^{-2}$ $h$/Mpc$)$ for the allowed neutrino masses in the $\nu w$CDM chains) in the matter power spectrum. Since quite large neutrino masses can be accommodated as we increase the number of PCs, this suppression becomes more marked with increasing PC number. Furthermore, on large scales (small $k$), there is an overall amplitude shift upwards, with the shift constituting a $>1 \sigma$ deviation from $\Lambda$CDM for the chains with 5 PCs. This is expected since the amplitude of the power spectrum increases as we increase the amount of dark energy in the universe (or decrease the amount of matter).

For the shear power spectrum, we show the fractional difference in the fourth source tomographic redshift bin $n(z_4)$ shown in Figure \ref{fig:nz} (it is nearly identical in all redshift bins so we chose to show a single one for clarity). The deviation away from $\Lambda$CDM of the shear power spectrum on large scales ($\ell \lesssim 100$) is larger than that of the matter power spectrum. Recall from Eq. \eqref{eq:plkappa} that the integral to calculate the shear power spectrum contains two distinct terms: the dimensionless matter power spectrum, and the lensing weight function, which encodes the background expansion. It is this additional dependence on the background which makes the difference with respect to $\Lambda$CDM larger in the shear power spectrum than the matter power spectrum, since as discussed above, neutrinos induce large differences in the background at high redshift.

The potentially large deviation from $\Lambda$CDM  suggests this observable as an exciting candidate to falsify $\Lambda$CDM, assuming weak lensing surveys can observe a wide enough $\ell$ range  to mitigate systematics such as the multiplicity bias, which is a nuisance parameter that shifts the overall amplitude of the cosmic shear signal. Upcoming data from the Wide Field Infrared Survey Telescope (WFIRST) \cite{2013arXiv1305.5425S,Schaan:2016ois}, the Large Scale Synoptic Survey (LSST) \cite{2009arXiv0912.0201L,Krause:2016jvl} and Euclid \cite{Amendola:2012ys}, will soon have better constraining power at large scales, although they will only be able to reach scales of about $\ell_{\rm min} \approx 30$. 

\bigskip

\section{Discussion}
\label{sec:discussion}

We have studied the constraints on the sum of the neutrino masses when marginalizing over principal components of the equation of state of a dynamical dark energy component that is allowed to cross the phantom barrier $w(z)=-1$. Exploring cosmological constraints on $M_{\nu}$ in the context of a general dark energy scenario is necessary because the background expansion is critical to probe the sum of the neutrino masses, meaning that there is a degeneracy between the dynamics of the dark sector and our ability to provide strong constraints on $M_{\nu}$. 

Our ability to constrain neutrinos with cosmology is further complicated by the fact that typical SN and (most) BAO measurements correspond to redshifts where dark energy cannot be ignored ($z \lesssim 0.7$). At higher redshifts, where dark energy is a subdominant component in the universe, the ability to compensate for neutrinos with large masses is diminished, and we found that this leads to observable deviations from $\Lambda$CDM. 

We investigated the effect of massive neutrinos on a variety of cosmological probes of geometry and growth within our cosmic-acceleration scenario and found that, by giving the dark energy equation of state more freedom than in traditionally-used parametrizations, dark energy can undo changes in the background expansion induced by the presence of massive neutrinos at late times. This was visible in the luminosity distance, where for $z < 2$ the posteriors for the $\Lambda$CDM and $\nu w$CDM chains agreed. 

This effect had two important consequences. First, much larger neutrino masses can be accommodated within the data: the upper bound of $M_{\nu}$ is as high as $0.38$ eV (95\% C.L., 3 PCs) or $ 0.55$ eV (95\% C.L., 5 PCs), when including weak lensing data. Second, at higher redshifts, where dark energy is a subdominant component in the universe and we do not have supporting data, there are large ($\sim 1\sigma$) deviations from $\Lambda$CDM in the background, since such large neutrino masses are not generally allowed by the combination of CMB and BAO data in $\Lambda$CDM (which is why analyses using combinations of these data sets find stringent upper bounds $M_{\nu} < 0.12$ eV). Furthermore, these large changes to the expansion history of the universe are also visible in the matter power spectrum at $z\approx0$, where there is a large amplitude increase on large scales in $\nu w$CDM with respect to $\Lambda$CDM and, consequently, on the shear power spectrum, since it is a measure of the integrated large-scale structure along the line-of-sight.

Since BAO measurements can probe amplitude shifts in distances, our results can be seen as a compelling case to pursue the BAO signal with high-$z$ tracers of the underlying baryonic density field ~\cite{2013A&A...552A..96B,2011A&A...534A.135L}. High-redshift Type IA supernovae can also be critical, since they can constrain changes in the shape of the luminosity distance induced by the transition between the low- and high-$z$ behavior. These measurements are within reach in the near future. For instance, recently there was a first detection of the BAO with the Lyman-$\alpha$ forest, at $z=2.4$ \cite{Slosar:2013fi}. Furthermore, experiments like eBOSS \cite{eBOSS} and DESI \cite{Aghamousa:2016zmz} will provide distance measurements at these redshifts in the very near future. 

The large amplitude offset in the shear power spectrum also makes it an exciting candidate to falsify $\Lambda$CDM, particularly since future surveys like LSST, Euclid, and WFIRST will reach very large scales, down to $\ell_{min} \approx 30$ (although this might not be enough to mitigate systematics that could hinder the use of this observable to falsify $\Lambda$CDM). As an aside, having more precise weak lensing information will be interesting due to the $S_8$ tension between \textit{Planck} and other weak lensing surveys, since to date there is no satisfying mechanism to solve it (although plenty of exotic models have been proposed); several weak lensing measurements \cite{2017MNRAS.465.1454H,Kohlinger:2017sxk,2017MNRAS.465.2033J,Troxel:2017xyo,Hikage:2018qbn} seem to have values of $S_8$ that are slightly lower than those of \textit{Planck}. 

Previous works have shown that allowing neutrino mass to vary when inferring cosmological parameters from weak lensing data sets lowers $S_8$: for instance, Ref. \cite{Hikage:2018qbn} showed that their best-fit value of $S_8$ was lowered with respect to $\Lambda$CDM by 0.5$\sigma$ when allowing neutrino mass to vary. Adding neutrinos to \textit{Planck} data has a similar effect, meaning that simply extending the base $\Lambda$CDM model by allowing $M_{\nu} \neq 0.06$ eV does not solve the tension.

Here we have seen that our PC-built equation of state further reduces $S_8$, and would have this effect if we considered a CMB likelihood and a WL likelihood independently, meaning that the additional freedom given to the dark energy component does not solve the tension either.

Note that, conversely, previous works that used non-parametric equations of state for dark energy were optimistic about the prospect of an evolving dark energy to solve cosmological tensions: Ref. \cite{Zhao:2017cud} found that they could mitigate the tensions in $H_0$ (between local $H_0$ and \textit{Planck}) and $\Omega_m$ (between BOSS and \textit{Planck}) with such an evolving dark energy model.

Our goal for this paper was to construct a model-independent $w(z)$ and see how the constraints on $M_{\nu}$ compared not only to $\Lambda$CDM but also to other works that have considered $M_{\nu}$ in the context of specific parametrizations for $w(z)$. As we have already mentioned, Ref. \cite{Vagnozzi:2018jhn} found $M_{\nu} < 0.19$ eV in their phantom, two-parameter dark energy cosmology. However, Ref. \cite{dePutter:2008bh} showed, by choosing specific parametrizations for $w(z)$ and projecting them onto the PC basis, that the first three PCs contain most of the information, which means that two-parameter models could be neglecting relevant information, leading to constraints on $M_{\nu}$ that are artificially tight. Here, we have seen that with only 3 PCs we have already opened up the allowed parameter space of neutrinos considerably ($M_{\nu} \lesssim 0.4$). 
This puts into question the claim that current data prefers $M_\nu \ll 0.3$ eV, since we have shown that this depends quite dramatically on our assumption about the behavior of dark energy.

However, this is a double-edged sword, and one must be careful when constructing a PC basis. As we point out throughout the text, the discrepancies between $\Lambda$CDM and $\nu w$CDM are significantly more apparent with 5 PCs than with 3. Looking at Figure \ref{fig:mnu_S8} it is clear that as we add more PCs to parametrize $w(z)$ the constraint on $M_{\nu}$ is loosened. Since most of the information is contained in the first few PCs, as we include higher PCs we are including modes that the SN cannot probe well. The higher PCs are essentially given free rein to alter the background cosmology, consequently opening up the parameter space of neutrinos even more. This means it is possible for theories of $w(z)$ to fit SN, but they increase $M_{\nu}$ posteriors considerably. One must therefore be cautious, and we leave results with 5 PCs as an extreme example. 

Ultimately,  there is no theoretical reason to favor the CPL (or similar) parametrizations over others, and models yet to be investigated might rely on higher PCs to be distinguished from $\Lambda$CDM~\cite{2010MortHuHuterer}, which is why pursing a model-independent approach to parametrizing $w(z)$ (and being cautious when building a basis so as to not have many unconstrained basis vectors) is an attractive alternative.

Unlike parametric forms for $w(z)$, using PCs allows us to remain agnostic with respect to what the alternative to the cosmological constant may be, since at present there is no strong theoretical or observational support for any particular exotic dark energy scenario. 
It is worth noting that the main caveat of the PCA method lies in the fact that it does not assess by itself the physical plausibility of the $w(z)$ shapes that are marginalized over. 

Finally, it is important to keep in mind that, as we mentioned in Section \ref{sec:data}, PCs oscillate quite drastically at ultra-low redshifts (outside the region of SNe support), but they do so by construction: they are unphased by physical assumptions, and are only driven by the data. Although many dark energy models are very smooth, there is no \textit{a priori} reason to believe that dark energy models with a low-redshift oscillatory $w(z)$ are unphysical (and in fact some models do predict such behavior \cite{2009PhRvD..80f7301M}). If one wanted to use the PCA method but had a strong reason to rule out the low-redshift oscillations allowed by the data, one could impose a prior that would punish such behavior, thus limiting the behavior of the PCs outside the redshift range supported by the data. 
In our implementation, where we have not imposed any prior at low-redshift to remain agnostic, the low-redshift oscillations degrade our ability to use local $H_0$ to constrain $M_{\nu}$. We are therefore not claiming that every model of dark energy would allow $M_\nu \sim 0.5$ eV, but merely that it is still possible to generate models that could make such extreme masses compatible with the data. 

\acknowledgments


We thank M. Gerbino, R. J. Foley, R. Hounsell, R. Kessler, E. Krause, J. Mu\~noz, D. Scolnic and D. Gates for helpful discussions. We also thank R. Kessler and W. Hu for providing us extensive computational resources granted by the University of Chicago Research Computing Center. CD and ADR were supported by NSF grant AST-1813694, Department of Energy (DOE) grant DE-SC0019018, and the Dean's Competitive Fund for Promising Scholarship at Harvard University. VM was supported by NASA ROSES ATP 16-ATP16-0084 and NASA ADAP 16-ADAP16-0116 grant. VM acknowledges the University of Arizona High Performance Computing center for providing computational resources. 

\newpage

\appendix

\section{Results for the \textit{Reduced} data set}\label{app:A}

We show the $1\sigma$ confidence levels for the posteriors of the luminosity distance and the Hubble expansion rate (Figure \ref{fig:Hz_DL_red}) and the matter power spectrum at $z=0$ (Figure \ref{fig:power_spectra_red}) for the \textit{Reduced} data set, in analogy with Figures \ref{fig:Hz_DL} and \ref{fig:power_spectra} in the main text for the \textit{All} data set. Note that we do not show the shear power spectrum since the \textit{Reduced} data set does not have weak lensing data.

\begin{widetext}

\begin{figure}[H]
\includegraphics[width=1.0\textwidth]{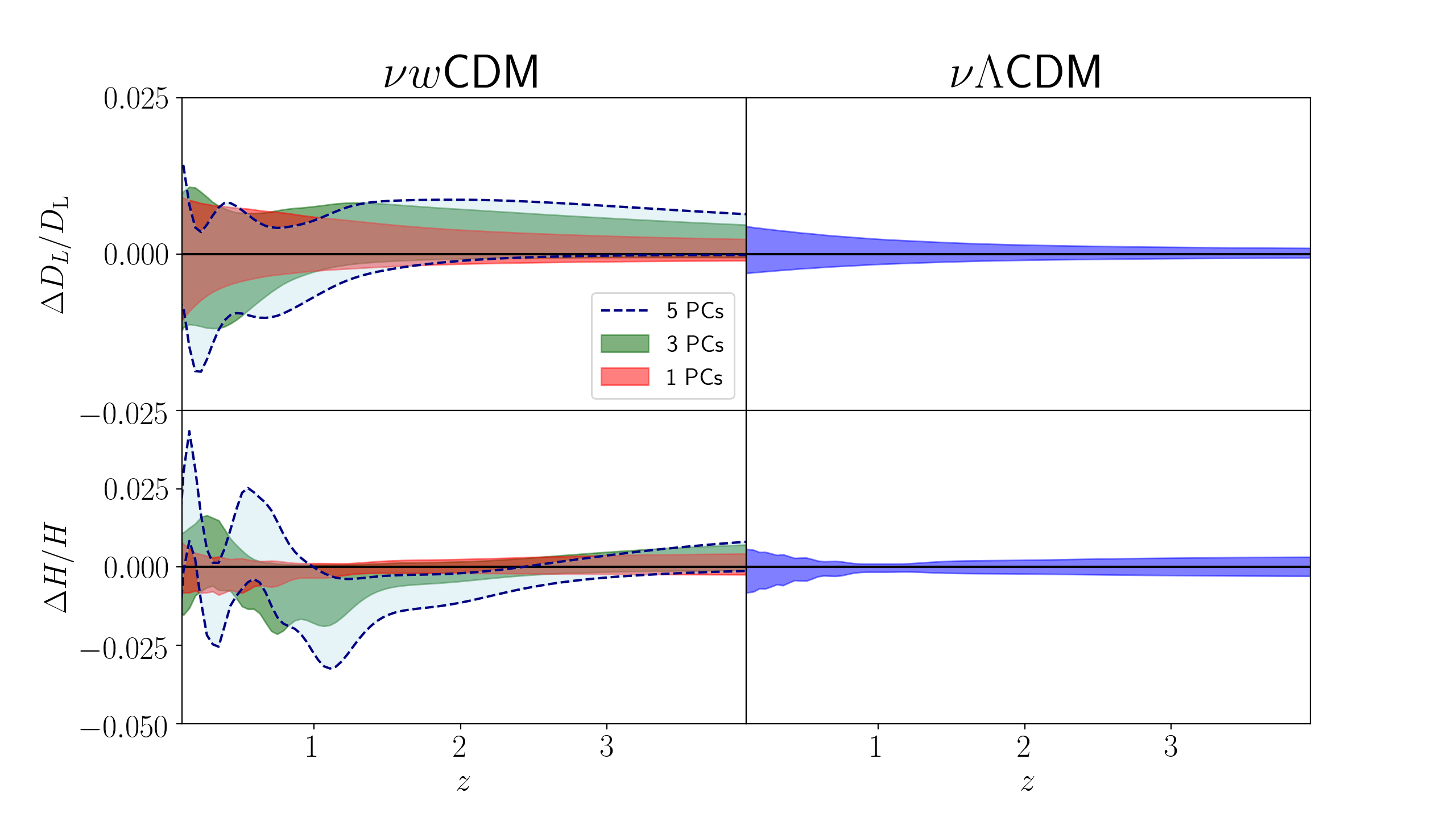}
\caption{Fractional difference for the luminosity distance $D_L(z)$ (top row) and the Hubble expansion rate as a function of redshift $H(z)$ (bottom row) for $\nu w$CDM (left) and $\nu\Lambda$CDM (right), with respect to $\Lambda$CDM, for the \textit{Reduced} data set. Bands correspond to $1\sigma$ confidence levels.
}\label{fig:Hz_DL_red}
\end{figure}

\begin{figure*}
\includegraphics[width=1.0\textwidth]{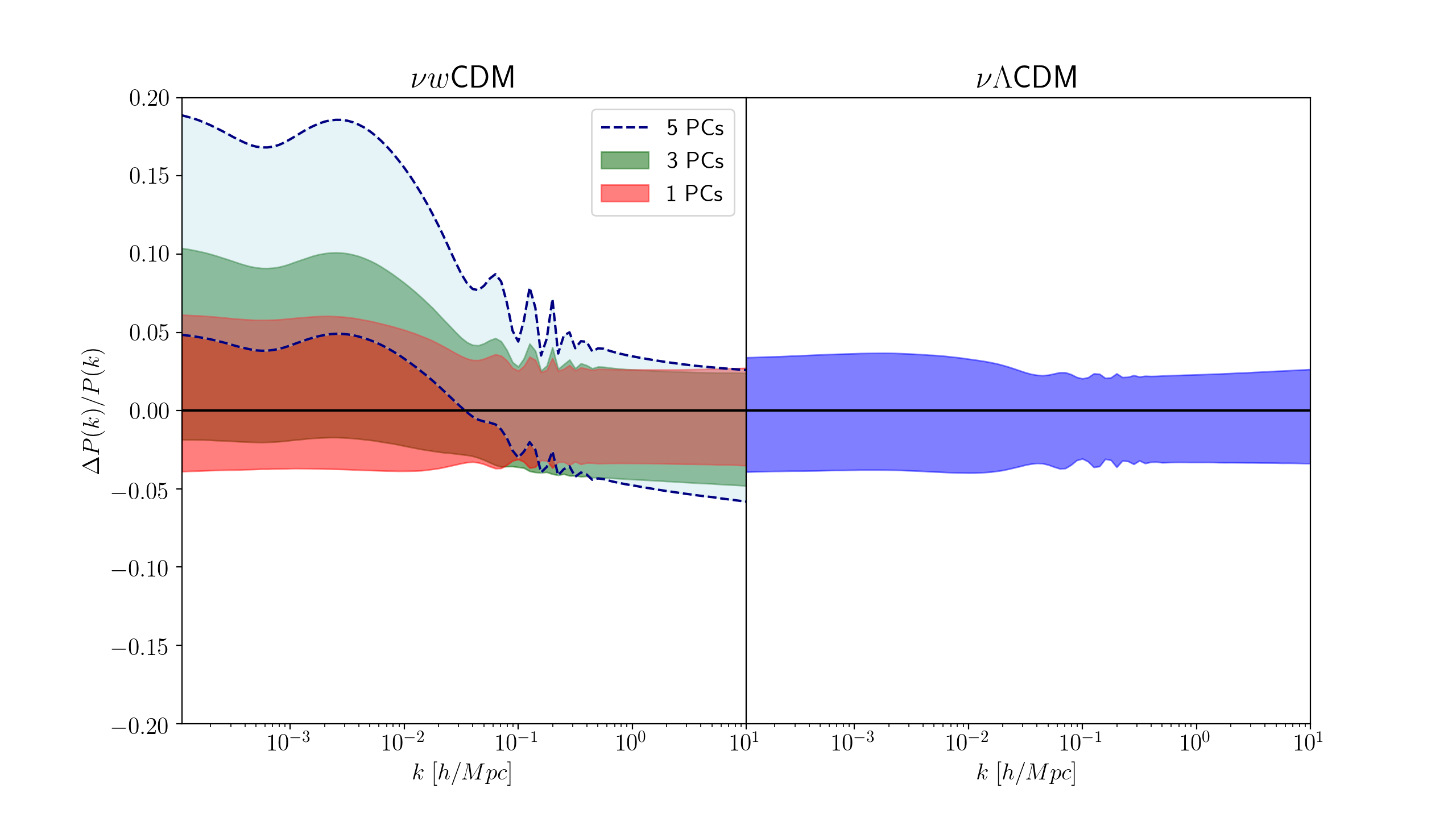}
\caption{Fractional difference for the matter power spectrum at $z=0$ $P(k)$ (top row) and the shear power spectrum $P_l^{\kappa}$ (bottom row) for $\nu w$CDM (left) and $\nu\Lambda$CDM (right) with respect to $\Lambda$CDM, for the \textit{Reduced} data set. Bands correspond to $1\sigma$ confidence levels.}\label{fig:power_spectra_red}
\end{figure*}

\end{widetext}

\newpage 


\bibliography{main}

\end{document}